# Histotripsy of blood clots within a hollow cylindrical transducer for aspiration thrombectomy applications

Li Gong[a,b,*], Alex R. Wright[b], Kullervo Hynynen[a,b,c], David E. Goertz[a,b]

a Department of Medical Biophysics, University of Toronto, Toronto, M5S 1A1, Canada
b Sunnybrook Research Institute, 2075 Bayview Avenue, M4N 3M5, Toronto, Canada
c Institute of Biomaterials and Biomedical Engineering, University of Toronto, Toronto, Canada

*Corresponding author at: Physical Sciences Platform, Sunnybrook Research Institute, Sunnybrook Health Sciences Centre, 2075 Bayview Avenue, Rm S665a, Toronto, ON, M4N 3M5, Canada. E-mail address: li.gong@mail.utoronto.ca (L. Gong).

**Abstract**

Thrombolytic occlusions in stroke, pulmonary embolism and the peripheral vasculature are increasingly treated with aspiration, a catheter-based approach that employs suction to extract clots through a hollow catheter lumen. Unfortunately, aspiration is frequently unsuccessful in extracting more challenging clots, which can become 'corked' in the distal tip. We hypothesize that clot extraction can be enhanced by using histotripsy to degrade the mechanical integrity of clot material within the lumen of a hollow cylindrical transducer which can be situated at the tip of an aspiration catheter. To demonstrate the feasibility of degrading clot material within the lumen of a hollow cylindrical transducer, the effect of pulsing schemes on the lesion generation within clots was assessed using a retracted clot model. A radially-polarized cylindrical transducer (2.5/3.3 mm inner/outer diameter, 2.5 mm length, PZT) working at 6.1 MHz was used to degrade retracted porcine clots with pulse lengths of 10, 20, and 100 µs, pulse repetition frequencies (PRF) of 100, 500, and 1000 Hz, using treatment times from 0.1 to 10 seconds (n = 5 clots/condition). 3D ultrasound scans and bisected optical examinations of treated clots confirmed the formation of liquified zones. Lesions could form within 0.1 seconds along the central axis of the transducer and then grow in diameter and length over time. The lesion volume was found to be highly dependent on the exposure scheme, with the largest lesion volume associated with the 10 µs pulse length 1000 kHz PRF case. Collectively these results demonstrate the feasibility of degrading blood clots within hollow cylindrical transducers, which suggests their potential for enhancing aspiration-based mechanical thrombectomy.

**Introduction**

The blockage of large blood vessels by clots is a significant cause of mortality and morbidity globally [1]. This can occur in various conditions such as ischemic stroke, myocardial infarction, and pulmonary embolism (PE), as well as in peripheral vessels (e.g. deep venous thrombosis, DVT) [2–5]. Catheter-based mechanical thrombectomy methods are increasingly used to treat acute thrombotic blockages [6]. One prominent approach, aspiration, involves the use of suction to extract clots through the lumen of a hollow catheter. While this technique has significantly improved patient care, aspiration attempts are often unsuccessful or require multiple passes, which can lead to less favourable patient outcomes [7–9]. A primary factor leading to unsuccessful aspiration is the inability of the achievable vacuum pressure to bring the clot fully into the catheter tip. For a clot to be fully ingested, it must undergo progressive deformation as it enters the tip, which leads to resistance that must be overcome by the vacuum. In many cases, such as for stiff clots or larger volume clots, 'corking' can occur where a portion of the clot material remains external to the catheter tip [10–17]. Consequently, there is significant interest in improving aspiration mechanical thrombectomy technology to facilitate the ingestion of clot material into the catheter.

The motivation for the present study is the development of a catheter-based ultrasound method for enhancing the ability of aspiration thrombectomy to extract challenging clots. The configuration of the catheter involves situating a hollow cylindrical transducer (HCT) within the tip of an aspiration catheter. Histotripsy will then be employed within the HCT lumen to break down the clot material as it enters into the catheter tip to reduce its mechanical resistance and thereby advance into the catheter more readily. Histotripsy is an established approach for using cavitation clouds to liquify tissue within the focal region of a transducer [18]. With few exceptions [19–21], histotripsy systems have employed large scale extracorporeal spherically focused transducers or arrays. A number of previous studies have demonstrated the use of histotripsy to degrade thrombus in-vitro [22–27] and in preclinical models [28–31].

A central challenge for the implementation of the proposed aspiration catheter approach is to achieve sufficiently high pressures within the lumen of radially polarized HCTs to induce cavitation. In recent work [32], we investigated the feasibility of generating cavitation within the lumens of intravascular scale (3.3/2.5 mm outer/inner diameter) HCTs. This was achieved by operating the transducers at a frequency within the thickness resonance mode bandwidth, where the resulting wall motion is predominantly associated with the generation of cylindrical waves. Axisymmetric cylindrical waves can be considered as the superposition of converging and diverging waves (to and from center respectively) [33], and in the case of an HCT there is the additional contribution of waves that are reflected at the lumen surface. By selecting an appropriate operating frequency and using sufficiently long pulses, standing wave constructive interference patterns can be generated. In these circumstances a high-pressure region concentrated along the central axis was shown to be present, along with side lobes of progressively decreasing amplitude toward the lumen wall. Using pulse lengths of 10 and 100 μs at 5 MHz for example, estimated pressures in excess of 20 MPa were generated, which were sufficient to produce cavitation clouds within degassed water.

The objective of this study is to investigate the feasibility of creating histotripsy lesions within clots that are situated in the lumen of an HCT. Retracted porcine clots were introduced into the HCT lumen prior to the commencement of exposures. A range of pulsing schemes were assessed to demine the effects of pulse length, pulse repetition frequency, and exposure duration. Lesion formation within the clots was monitored under high frequency ultrasound imaging (40 MHz), and treated clots were evaluated with high-resolution 3D ultrasound scans to quantify the lesion volume and geometry. Clots were then bisected to further validate lesion formation optically.

**Methods**

**Transducer configuration**

The transducer was of the same design as described in [32], with an element length of 2.5 mm, outer diameter of 3.3 mm, inner diameter of 2.5 mm, and material type of DL-47 (DeL Piezo Specialities, FL, USA). Micro-coaxial wire was attached to the inner surface signal electrode and outer surface ground electrode using silver epoxy (Epotek H20E, Epoxy Technology Inc, MA, USA), and a PTFE catheter liner (PTFE Sub-Lite-Wall™ Liner 0.095" ID 0.0015" wall thickness, Zeus Company Inc, SC, USA) with an outer diameter closely matching the inner diameter of the transducer was attached with epoxy (Epotek 301, Epoxy Technology Inc, MA, USA) to the inner lumen of the cylinder, and trimmed flush with the front face of the element. The liner extended backwards 3 cm before connecting to rigid tubing that was attached to a 60 mL syringe at the proximal end. The impedance of the transducer was measured in water from 300 Hz to 8 MHz in 5 kHz steps using a network analyser (AA-30.ZERO, RigExpert, Kyiv, Ukraine). To determine the operating frequency, a frequency sweep at an applied peak voltage of 65 V (as measured across the transducer) with 10 µs pulses and a 10 Hz PRF was performed in the range of 4 to 7 MHz with 10 kHz steps. The frequency of 6.1 MHz showed the highest visual level of cavitation activity on B-mode imaging with water in the lumen and was chosen as the treatment frequency. This frequency corresponded to a minimum on the impedance curve, consistent with the results of [32]. The pressure profile within the HCT operating at this frequency was simulated using the finite element analysis (FEA) software package OnScale™, according to the procedures described in [32]. A high-pressure lobe is present along the main transducer axis, along with side-lobes that decay towards the inner lumen wall (Figure 1).

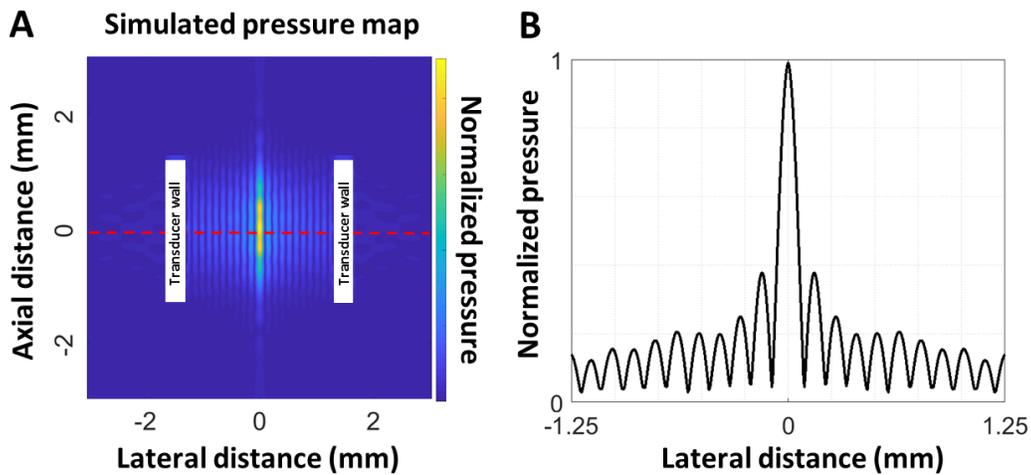

*Figure 1: The pressure map in water (A) and pressure distribution on the center lateral lines (B) at 6.1 MHz from simulation. Pressures are normalized to the peak of the pressure map.*

**Blood clot preparation**

Clots were prepared according to previous sonothrombolysis studies [24,25,27,34,35]. Blood was collected from the femoral vein of pigs weighing approximately 70 kg into 3.2% sodium citrate vacutainers (BD Vacutainer™ Plastic Blood Collection Tubes with Sodium Citrate, Thermo Fisher

Scientific Inc, MA, USA) according to institutional animal ethics standards. Clots were prepared within 1 hour of blood collection by mixing the anti-coagulated blood with 100 mmol/L calcium chloride (200 μL of $CaCl_2$ per 1 mL of anti-coagulated blood) and then transferring to 2 mL borosilicate pipettes (Fisherbrand™ Disposable Borosilicate Glass Pasteur Pipets, Thermo Fisher Scientific Inc, MA, USA). The pipettes were sealed with parafilm and incubated in a water bath at 37 °C for 3 hours before being stored at 4 °C for 3 days. The retracted clots (3~4 mm diameter) were then extracted from the pipettes and cut into homogenous 5 mm long sections immediately prior to experiments.

**Experimental configuration**

The experimental configuration is shown in Figure 2. A VisualSonics Vevo 2100 ultrasound system with a 40 MHz probe (MS550D) was used for B-mode imaging during treatment. The imaging plane of the probe was aligned along the central axis of the annular transducer, and positioned 4 mm away from the distal face of the transducer. For each treatment, 10,000 frames were recorded at a frame rate of 100 Hz. The imaging system was synchronised with treatment pulses using a pulse delay generator (Model 575, Berkeley Nucleonics Corp, CA, USA) such that axial interference bands associated with the treatment sonication pulses were positioned in the same lateral location for each recording. The pulse delay generator triggered a function generator (AFG3102, Tektronix, OR, USA) whose signals were amplified (A150, E&I, NY, USA) before being sent to the transducer. Reported applied voltages were measured across the transducer using a 100:1 voltage probe. Experiments were performed in a tank of deionized water held at 30 °C. Clots were aspirated into the transducers using the 60 mL syringe, such that the 5 mm length clots were positioned roughly centered within the 2.5 mm length transducer. After treatment, the clots were immediately removed from the transducer using light pressure from the syringe and captured into a dish containing PBS. The clots were then 3D scanned (0.05 mm plane spacing) with a second Vevo 2100 system using a 50 MHz probe (MS700), typically this was done within 2-3 minutes after exposures ended. After 3D imaging, clots were either bisected and the internal lesions were imaged under an optical microscope at 2.5X magnification, or the intact clots were fixed in 10% formalin and sliced at 5 μm thickness sections before H&E staining for histological analysis.

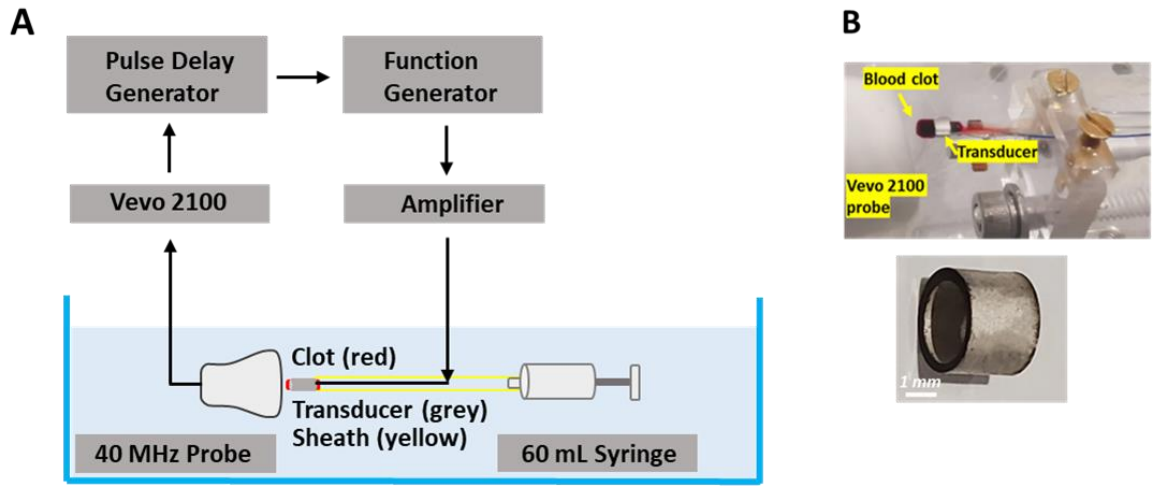

*Figure 2. (A) Schematic of experimental configuration. (B) Images of experimental setup with clot inside transducer (top) and closeup view of raw element used in transducer (bottom). The 2.5 mm length element has a 2.5 mm inner diameter and 3.3 mm outer diameter.*

**Histotripsy pulsing scheme**

Pulse lengths of 10, 20 and 100 µs were assessed, which spanned the range employed in [16] and were sufficiently long to result in standing waves within the lumen. PRFs of 100, 500, and 1000 Hz were employed, which resulted in duty factors ranging from 0.001 to 0.01. Treatment durations ranging from 0.1 to 10 seconds were assessed. Excitation voltage was fixed at 82 V (peak) for all treatments. Supplementary Table 1 shows the 18 pulsing schemes tested. 5 clots were treated per exposure condition. For each pulsing scheme, 3 to 5 clots were bisected for optical imaging, and for a subset of pulsing schemes 1 to 2 clots were fixed for H&E stained histology.

**Lesion quantification**

Lesions were quantified using 3D scanned B-mode images. The boundary of lesion was identified in a series of transverse slices through the scan and then obtain its volume. Each transverse frame showing a visible dark lesion was contoured manually, giving a total lesion volume. The lesion diameter and length were measured from the major and minor axis of the lesion on the sagittal slice showing the largest lesion area. For normalizing measurements relative to ultrasound exposure time, a measure of lesion volume change per pulse was calculated by dividing the measured lesion volume by total number of pulses for a given pulsing scheme.

Statistical significance of lesion measurements between different pulsing groups were tested using a one-way ANOVA (post-hoc: Tukey's) when comparing three or more groups or an unpaired t-test (two-tailed) when comparing two groups. Differences between groups were deemed significant when p<0.05. All statistical calculation were performed using Graphpad Prism. Results are presented as mean +/- standard deviation (SD). Statistical significance group comparisons and fold differences are shown in Supplementary Tables 2 to 6.

**Results**

Visible cavitation clouds as well as post-treatment lesions were observed on B-mode imaging for every pulsing scheme tested. Figure 3 shows frames from before, during, and after 10 seconds of treatment with 10 µs pulses and a 100 Hz PRF. In this representative example, a cavitation cloud is apparent from the first treatment pulse (0.01 second into treatment), and the size of the cavitation cloud grows over the duration of the treatment. Images post-treatment show hypoechoic lesions formed over the same areas as the cavitation cloud towards the end of treatments.

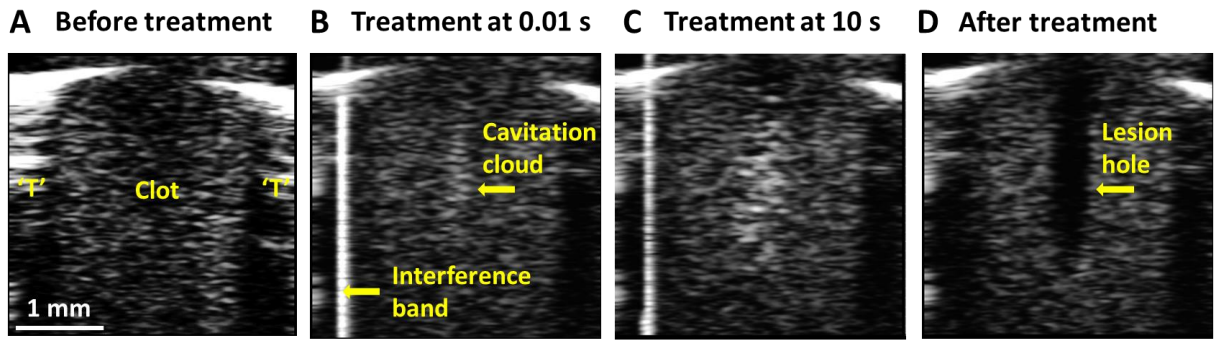

*Figure 3: Representative B-mode imaging showing the cross section of a clot situated in the transducer before, during, and after treatment. 'T' represents transducer walls. Treatment parameters were 10 µs pulse length, 100 Hz PRF, and 10 seconds treatment duration. Before treatment, the clot is visible as uniform speckle bounded by the side walls of the transducer. During treatment, the cavitation cloud is visible as a hyperechoic region along the central axis of the transducer. Immediately post treatment, the yellow arrow points towards the hypoechoic region visible over the same area of the cavitation cloud, indicating the generation of a lesion.*

An example of a 3D scan of a treated clot is shown in Figure 4. The multiplanar reconstruction shows that the hypoechoic lesion is located symmetrically within the center of the clot.

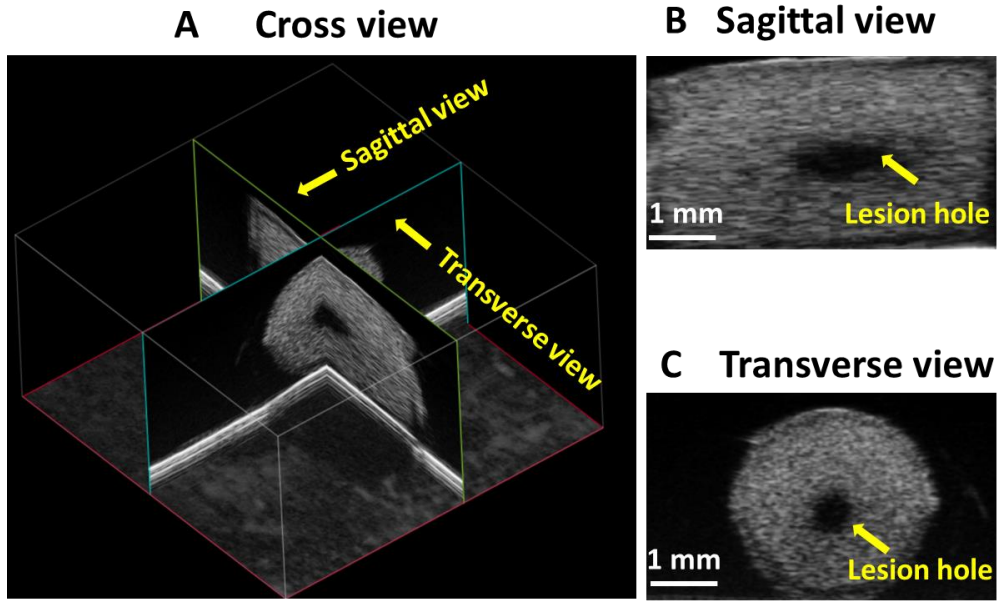

*Figure 4: (A) Representative 3D scanned B-mode images of a treated clot. (B) Sagittal view of a central slice along the length of the clot with a hypoechoic lesion situated in the center. The lesion is positioned along the length of the clot directly between where the faces of the transducer were situated. (C) Transverse view showing a cross section of the lesion hole centered along the axis of the clot. The clot was treated for 10 seconds with 10 μs pulse length and 100 Hz PRF.*

The stained slice of a treated clot in Figure 5A shows a lesion devoid of tissue bounded by the clot material. The detailed view of the lesion boundary in Figure 5B shows the transition region between the lesion and untreated clot.

Representative B-mode images and photographs of the treated clots are shown in Figure 6 for treatments with 10 μs pulses at a PRF of 1000 Hz as a function of treatment duration. Visible lesions are formed in as little as 0.1 seconds, and lesion size increases as a function of treatment duration. In both B-mode images and photographs the lesion boundary is generally well defined, although the transition zone between the lesion void and untreated clot is relatively larger for shorter treatments due to the smaller lesion size. At longer treatment durations for the 1 kHz case (but not the 100 Hz case), residual bubbles were frequently evident on the B-scan images as bright spots at the lesion boundary.

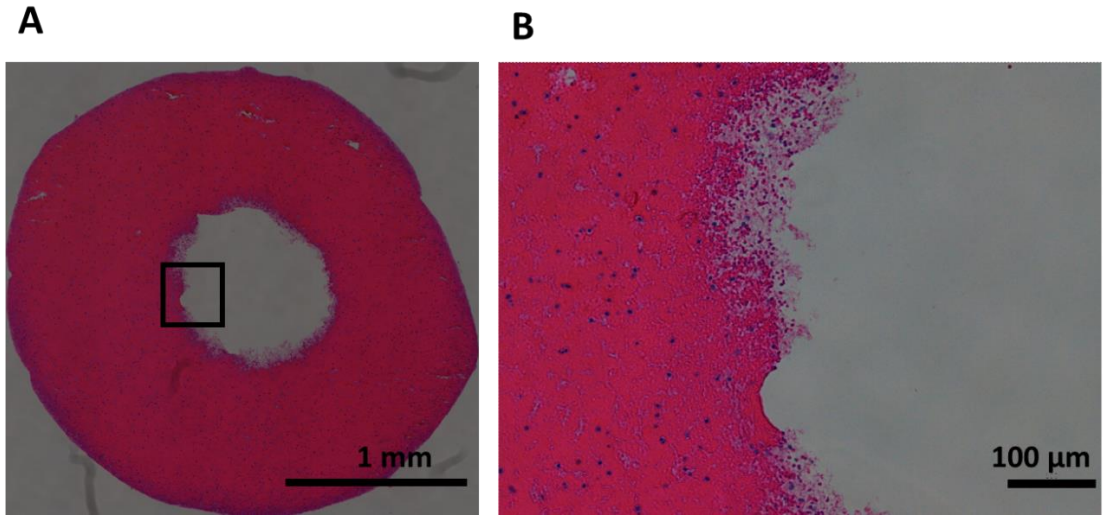

*Figure 5: (A) H&E stained histology slice of a treated clot. Central region shows lesion area devoid of tissue (B) Detailed view of lesion boundary. The clot underwent a 10 second treatments with 10 µs pulse length and 100 Hz PRF.*

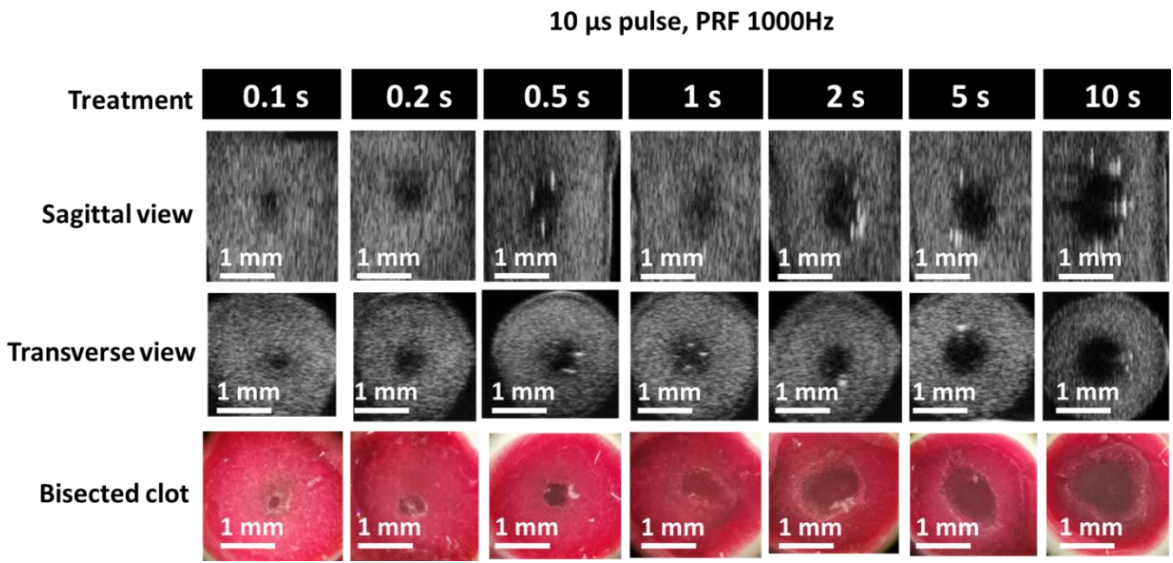

*Figure 6: Representative B-mode images (whole clots) and photographs (bisected) of treated clots for a pulse length of 10 µs and PRFs of 1000 Hz (0.01 duty factor). B-mode images are the sagittal and transverse slices with the largest lesion volumes taken from 3D B-mode scans. Optical images are from bisected clots imaged under a microscope at 2.5X magnification.*

For treatments with 10 µs pulse lengths, treatment durations of 1, 5 and 10 seconds were tested with a PRF of 100 Hz, while treatment durations of 0.1, 0.2, 0.5, 1, 2, 5, and 10 seconds were tested with a PRF of 1000 Hz, and the quantified lesion dimensions as a function of treatment duration for these treatments are summarized in Figure 7. For the 1000 Hz PRF case, the measurements confirm that lesion volume, diameter, and length increase with increasing treatment duration. The greatest relative increases in lesion volume occurs between 0.1 and 0.2 seconds, and growth continues until 10 seconds, albeit at a lower rate. In contrast, for the PRF of 100 Hz, there is no significant lesion growth between 1 to 10 seconds.

Comparisons of lesion dimensions as a function of PRF for different treatment durations at a pulse length of 10 µs are shown in Figure 8A. Generally, an increased PRF is associated with larger lesion diameters and volumes, although this trend appears more prominent as treatment durations increase. There is a significant difference in both lesion volumes and diameters between all PRFs tested for both 5 second (100 and 1000 Hz) and 10 second treatments (100, 500, and 1000 Hz). Lesion volumes increase with increasing PRF for 1 second treatments but the differences are not significant ($P>0.05$). The only other significant differences appear in lesion diameter for 1 second treatments when comparing 100 and 500 Hz PRFs, and for lesion lengths for 10 second treatments when comparing 100 and 1000 Hz and 500 and 1000 Hz. To gain insight into lesion formation efficiency, the change in lesion volume per pulse as a function of PRF and treatment duration are shown in Figures 8B and 8C respectively. The change in lesion volume per pulse decreases with increasing PRF and increasing treatment duration. For both 1 and 10 second treatments, there are significant decreases between 100 and 500 Hz and between 100 and 1000 Hz, and for 5 second treatments there is a significant decrease between 100 and 1000 Hz treatments.

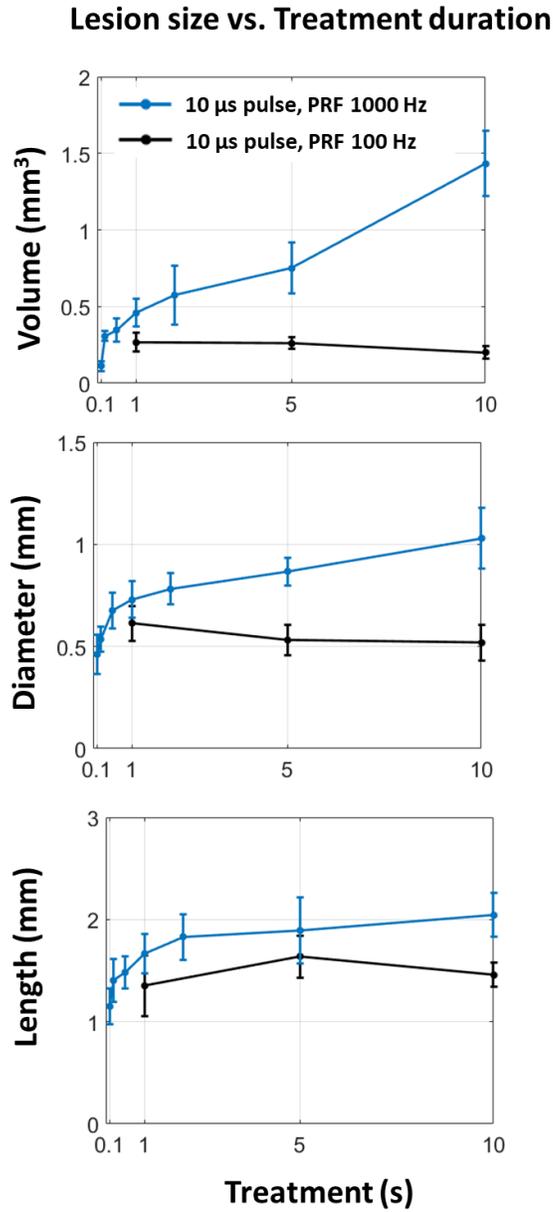

*Figure 7: Lesion measurements (volume, diameter and length) as a function of treatment duration for a pulse length of 10 µs and PRFs of 100 Hz (0.001 duty factor) and 1000 Hz (0.01 duty factor).*

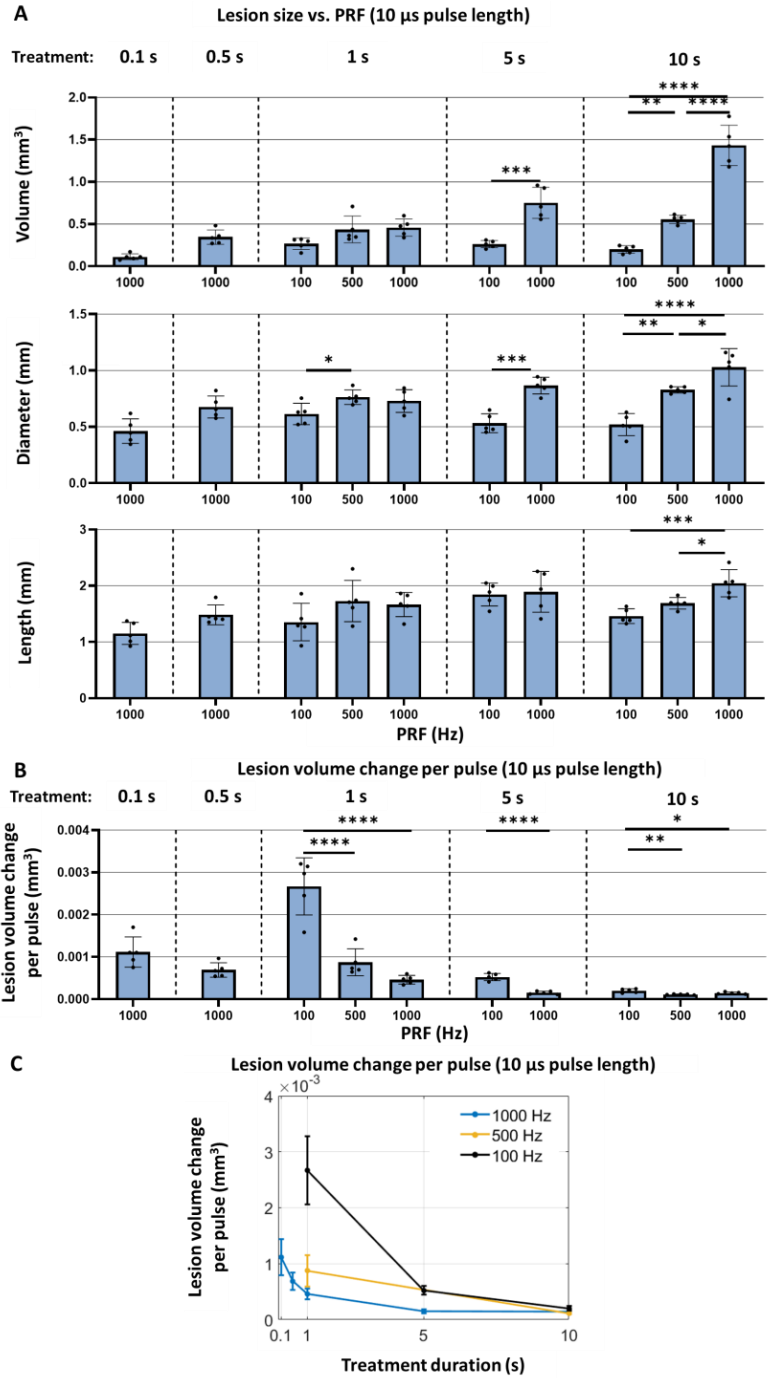

*Figure 8: (A) Lesion measurements (volume, diameter and length) as a function of PRF for treatment durations of 0.1 to 10 seconds. (B) Lesion volume change per pulse as a function of PRF for different treatment durations. (C) Lesion volume change per pulse as a function of treatment duration with 100, 500 and 1000 Hz PRFs. All pulsing schemes have a pulse length of 10 μs.*

A comparison of lesion dimensions as a function of pulse length for treatments of 1 and 10 seconds in duration and PRFs of 100, 500 and 1000 Hz can be seen in Figure 9A.

For a 1 second treatment at a PRF of 100 Hz, lesion volume, diameter, and length increase as a function of pulse length, although the increase in volume and diameter is only significant ($P<0.0001$ and $P<0.01$) between 20 and 100 µs pulse lengths and 10 µs and 100 µs pulse lengths, but not between 10 and 20 µs pulse lengths. At a PRF of 500 Hz, the lesion diameter and length increase ($P>0.05$) from 10 to 20 µs pulse lengths while the lesion volume slightly decreases ($P>0.05$). This inconsistency is due to the lesion diameter and length being measured on the sagittal slice with the largest area, while the lesion shape as a whole may be non-uniform. Differences in lesion diameter are only significant at 100 Hz, while no differences in lesion lengths at any PRF are significant.

For treatment lengths of 10 seconds, at 100 Hz the lesion volume, diameter, and length increase as a function of pulse length, although again the increase in lesion volume is more apparent from 20 to 100 µs ($P<0.0001$) compared to 10 and 20 µs ($P>0.05$). At 500 Hz, lesion volume and diameter increase non-significantly ($P>0.05$) between 10 to 20 µs pulse lengths, while average lesion length slightly decreases ($P>0.05$). Comparing lesion dimensions for a given pulse length across different PRFs, increasing PRF generally results in increased lesion dimensions. A subset of this data compared in terms of total exposure time can be seen in Supplementary Figure 1.

Figures 9B–D show lesion volume change per pulse for different treatment durations as a function of PRF and pulse length. In general, 100 µs pulse lengths show a greater lesion volume change than the 10 and 20 µs pulse lengths for all treatment durations, although not proportionally with the increased duration of ultrasound exposure. The 10 and 20 µs pulse lengths are similar to each other for 0–10 seconds and 0–1 seconds of treatment, while the 20 µs pulse length shows greater volume change at 500 Hz when looking at 1–10 seconds of treatment. The change in lesion volume per pulse is much greater for all cases in the first 0–1 seconds of treatment than for 0–10 seconds or 1–10 seconds of treatment. Change in lesion volume per pulse is invariant with PRF for 0–10 second treatment, decreasing with PRF between 0–1 seconds of treatment, and increasing with PRF for 1–10 seconds of treatment.

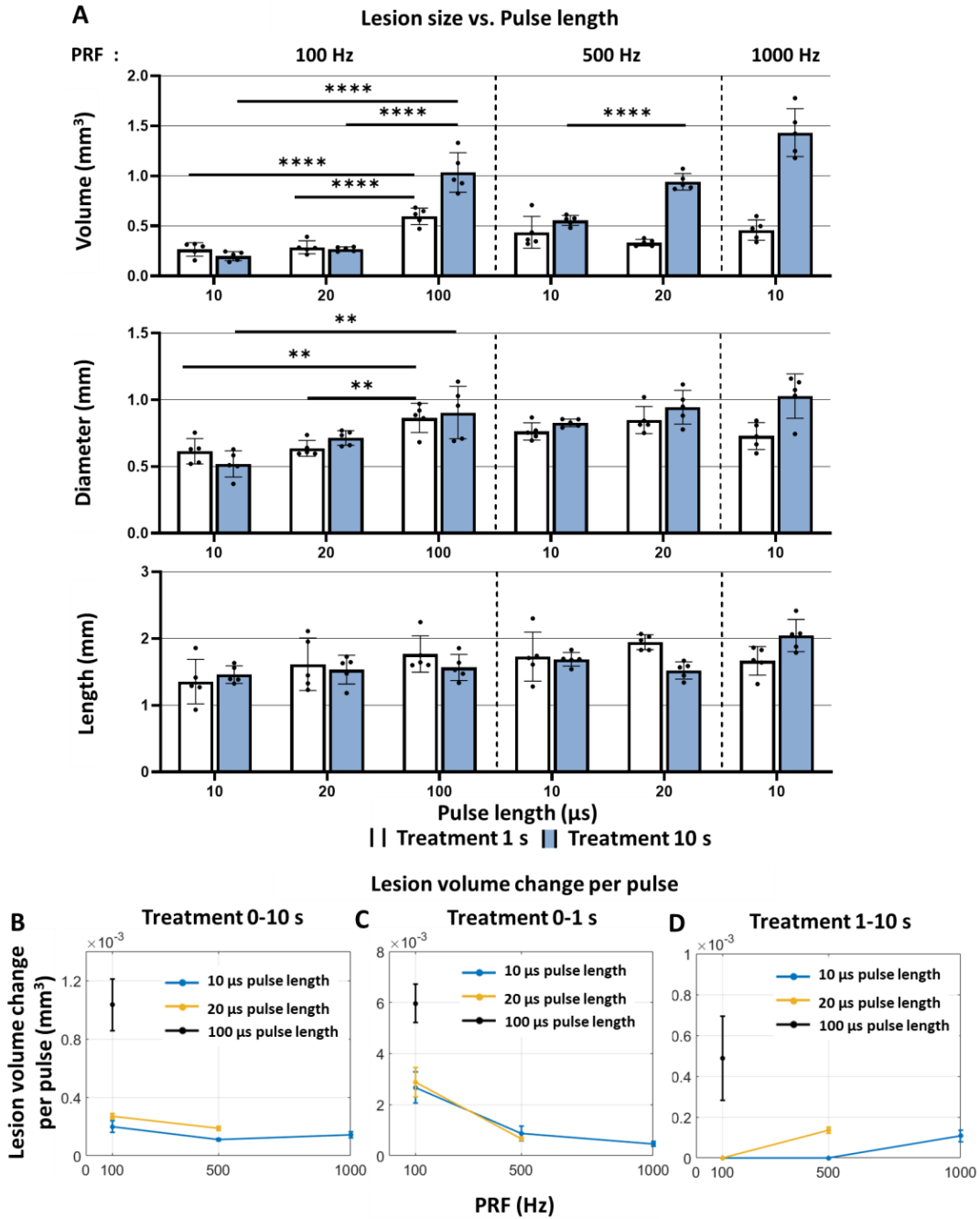

*Figure 9: Lesion measurements (volume, diameter and length), (A) as a function of pulse length for treatment durations of 1 and 10 seconds with PRFs of 100, 500 Hz and 1000 Hz. Lesion volume change per pulse as a function of PRF for different pulse lengths and treatment durations of (B) 0–10 seconds, (C) 0–1 seconds, and (D) 1–10 seconds, obtained by subtracting the average volume for 1 second duration treatments from the average volume of 10 second duration treatments and dividing by the net number of pulses over 9 seconds.*

**Discussion**

In previous work it was established that cavitation could be generated within the lumen of an intravascular scale HCT transducer [32]. The results of the present study have demonstrated the feasibility of creating lesions in clots situated within the lumen of an HCT. Lesion formation was found to begin in as little as 0.1 second along the central transducer axis and could expand radially and in length with increasing time. The eroded volume was highly dependant on the pulsing parameters employed.

It is useful to consider these results in the context of histotripsy that is performed with spherically focused transducers, where the influence of pulse length, PRF and transducer focal characteristics have been well investigated [19,22,36,37]. These studies have typically employed high gain transducers that resulted in substantially nonlinear pulses. Histotripsy is often classified based on the pulse characteristics and the underlying physical processes involved in cloud formation. Intrinsic histotripsy relies on generating cavitation clouds from endogenous cavitation nuclei using highly nonlinear pulses lasting 1–2 cycles. With this method, bubble expansion occurs during the initial rarefactional phase, resulting in a relatively small cloud in the central focal area. For shock scattering, pulse durations typically range from 5 to 20 cycles, where a cloud forms during the early phase of the pulse, and subsequent cycles are reflected and scattered by the initial cloud, amplifying its growth, which can propagate towards the transducer. Boiling histotripsy employs pulses lasting approximately 1–100 milliseconds. The longer duration, combined with absorption effects, leads to temperature increases that reduce the pressure threshold necessary for cavitation cloud formation. Pulse lengths intermediate between shock scattering and boiling histotripsy have also been employed, which have recently been referred to as 'hybrid' [38–41]. Specific to thrombolysis, previous histotripsy studies have employed pulsing schemes in the intrinsic [23,29], shock scattering [26,28,42], hybrid (0.1 milliseconds) [30,43] and boiling [39,44] regimes.

The 10 µs pulses employed here lie within the shock scattering classification, though the degree of pulse nonlinearity remains to be established. The 100 µs pulses lie in the lower range of hybrid pulse lengths. Beyond pulse length similarities to previous work, it must also be considered that there are important differences between the pressure fields of spherically focused transducers and HCTs operating at a thickness mode, which may impact the development and evolution of cavitation clouds and lesions. In particular, in the absence of cavitation (or bubbles), the luminal lateral pressure profile (Figure 1) arises not just from the inherent cylindrical wave focal gain, but also from the interference between the inward and outward propagating waves along with reflections from the luminal wall [32]. Once a cavitation cloud is initiated, or if (residual) bubbles are present at the outset, it can be expected that it will reduce the constructive interference component of the field. Further, additional reflections from the cloud boundary and emissions from the cloud will impact the resulting pressure field. This is a complex process which has yet to be investigated, but the factors involved should be considered in interpreting the results of the present study.

The 10 µs pulse results show an early-stage initiation (< 1 second) of bubble clouds and lesions along the transducer axis, broadly corresponding to the location of the high-pressure levels of the 'baseline' beam (no bubbles present). A comparison of the 0.1 seconds treatment duration at 1 kHz and 1 second treatment duration at 100 Hz time points, both of which follow 10 transmit pulses, indicates that the 100 Hz is more efficient at lesion generation than 1 kHz, on a per pulse basis (2.4 fold). This result is consistent with what has been observed with spherically focussed transducers in a shock scattering pulse regime, which is attributed to the effects of bubble 'memory' [45,46]. That is, if the time interval between pulses is sufficiently low (PRF sufficiently high) then residual bubbles from a given pulse will still be present when the ensuing pulse arrives. This will impact the nature of the bubble clouds that form (density, bubble size and location), one aspect of which is to increase the spatial correlation of bubble sites within the cloud which will in turn reduce the tissue erosion rate and therefore lesion size [46]. An additional factor that may be occurring in the HCT case is the possible impact of residual bubbles on the pressure field that develops.

With increasing treatment duration (1–10 seconds) the 100 Hz case lesions did not grow, whereas 1 kHz lesion volumes continued to increase while expanding progressively radially outward beyond the on-axis baseline high pressure region. Indeed, by 10 seconds the lesion diameters were on the order of 1 mm, which extends well into the lower pressure regions of the baseline pressure profile (Figure 1). The reasons for this remain to be established. Aside from the higher number of pulses sent in the 1 kHz case, it must also be considered that once clouds (and lesions) are initiated along the central axis, the pressure field outside of the lesions will depart from the baseline pattern due to reflections, attenuation and emissions. It is possible that the effects of bubble memory may create a more favourable lateral (away from main axis) pressure field for erosion, through differences in cloud characteristics (e.g. density or bubble size impacting reflections). Separately, the presence of residual bubbles may facilitate lateral erosion at the lower pressure levels. Related to this, it was frequently observed that bubbles (bright spots on US imaging) were present at the lesion periphery in the post-treatment 3D ultrasound scans for the 1 kHz case, but not the 100 Hz case.

In terms of pulse length, erosion performance at 100 Hz for 10 and 20 µs were similar but the 100 µs performed better (3.8–5.2 fold) albeit at a higher duty factor (0.001 and 0.002 vs. 0.01 respectively). At 500 Hz, the 20 µs pulse performed better (1.7 fold) in the 10 second treatments, though with double the duty factor. For a 0.01 duty factor, the 10 µs pulse performed better than the 20 and 100 µs pulses in terms of erosion volume over 10 seconds (Supplementary Figure 2). As a point of reference, duty factors employed in previously studies using shock scattering and hybrid pulsing schemes have varied widely, but many are within a 0.001–0.01 range [28,30,43,44]. Aside from performance considerations, one consideration for limiting duty factors to ~0.01 or below is to limit the potential for thermal effects.

Collectively, the results indicate that lesion formation can be readily achieved over a wide range of pulsing conditions, and that the degree of erosion is highly dependant on the particular conditions employed. As this was a feasibility study, the focus was on using basic pulsing schemes- i.e. sending a series of pulses at a fixed PRF. The investigation of a wider range of pulsing schemes is also warranted in future work [47]. Importantly, an improved physical understanding of cavitation and histotripsy in the setting of intravascular scale HCTs is also necessary and would aid in the rational design of exposure approaches.

Ultimately the purpose of the approach investigated here is to employ aspiration catheter tip based HCTs to facilitate the ingestion of clots into the catheter. Under vacuum, clots must deform to enter into and pass through the tip and then proceed proximally within the catheter lumen. The results have shown that the (liquid filled) lesions created are centered within the clot, with a diameter that is dependant on the exposure conditions and duration. The 10 seconds time scale was selected to be relevant to aspiration procedures, which depending on the application can be tens of seconds or minutes long [48,49]. A central question that was not addressed here, and is the subject of future work, is if creating liquified zones within the clots will facilitate deformation and therefore enhance ingestion.

The HCT size employed here is the same as used in [32] to demonstrate the feasibility of generating cavitation within an HCT lumen. Its outer diameter is compatible with incorporating into a 11Fr catheter, which is on the larger range of aspiration catheters [50–52]. It should be noted that the proof of concept evaluated here is in principle compatible with being adapted to a wide range of HCT sizes, and therefore application areas, such as PE, DVT and stroke.

It should be noted that other catheter-based ultrasound approaches for performing thrombolysis in the setting of large vessel occlusions have been reported. The EKOS catheter is a clinically employed method for large vessel (DVT [53,54], PE [55]) sonothrombolysis. This catheter employs a series of transducers and side-ports situated along its distal portion. After fully inserting in into a clot, thrombolytic enzymes are infused and their activity is enhanced via low intensity ultrasound exposures, typically over 6–18 hours. Ultrasound catheters for cavitation mediated thrombolysis are also under development. These involve situating an ultrasound transducer at the catheter tip, such that the beam is forward looking, and releasing cavitation agents (microbubbles, droplets) and possibly thrombolytic enzymes in the vicinity of the catheter tip. The concept is to expose the clot boundary to cavitation, and thereby progressively erode it as the catheter is advanced. All approaches to date therefore involve the in-situ erosion of clot material to restore blood flow. In contrast, the approach proposed here involves enhancing the performance of a widely employed existing clinical approach to extract clot material.

**Acknowledgements**

This work was funded by NSERC Discovery and CIHR Project grants. The authors thanks Dr. Emmanuel Cherin for his guidance in using the Vevo imaging system; Sharshi Bulner for his assistance for histology imaging; Jennifer Barry, Shawna Rideout, Dallan McMahon and Stephanie Furdas for performing animal blood collection.

[49] A.K. Sista, J.M. Horowitz, V.F. Tapson, M. Rosenberg, M.D. Elder, B.J. Schiro, S. Dohad, N.E. Amoroso, D.J. Dexter, C.T. Loh, D.A. Leung, B.K. Bieneman, P.E. Perkowski, M.L. Chuang, J.F. Benenati, E.-P. Investigators, Indigo Aspiration System for Treatment of Pulmonary Embolism Results of the EXTRACT-PE Trial, Jacc Cardiovasc Interventions 14 (2021) 319–329. https://doi.org/10.1016/j.jcin.2020.09.053.

[50] R. Loffroy, N. Falvo, K. Guillen, C. Galland, X. Baudot, E. Demaistre, L. Fréchier, F. Ledan, M. Midulla, O. Chevallier, Single-Session Percutaneous Mechanical Thrombectomy Using the Aspirex®S Device Plus Stenting for Acute Iliofemoral Deep Vein Thrombosis: Safety, Efficacy, and Mid-Term Outcomes, Diagnostics 10 (2020) 544. https://doi.org/10.3390/diagnostics10080544.

[51] B. Robertson, E. Neville, A. Muck, M. Broering, A. Kulwicki, B. Kuhn, M. Recht, P. Muck, Technical success and short-term results from mechanical thrombectomy for lower extremity iliofemoral deep vein thrombosis using a computer aided mechanical aspiration thrombectomy device, J. Vasc. Surg.: Venous Lymphat. Disord. 10 (2022) 594–601. https://doi.org/10.1016/j.jvsv.2021.11.002.

[52] R. Sedhom, P. Abdelmaseeh, M. Haroun, M. Megaly, M.A. Narayanan, M. Syed, A.M. Ambrosia, S. Kalra, J.C. George, W.A. Jaber, Complications of Penumbra Indigo Aspiration Device in Pulmonary Embolism: Insights From MAUDE Database, Cardiovasc. Revascularization Med. 39 (2022) 97–100. https://doi.org/10.1016/j.carrev.2021.10.009.

[53] C. Owens, Ultrasound-Enhanced Thrombolysis: EKOS EndoWave Infusion Catheter System, Semin Intervent Radiol 25 (2008) 037–041. https://doi.org/10.1055/s-2008-1052304.

[54] M. Dumantepe, I.A. Tarhan, A. Ozler, Treatment of Chronic Deep Vein Thrombosis Using Ultrasound Accelerated Catheter-directed Thrombolysis, Eur. J. Vasc. Endovasc. Surg. 46 (2013) 366–371. https://doi.org/10.1016/j.ejvs.2013.05.019.

[55] G. Piazza, B. Hohlfelder, M.R. Jaff, K. Ouriel, T.C. Engelhardt, K.M. Sterling, N.J. Jones, J.C. Gurley, R. Bhatheja, R.J. Kennedy, N. Goswami, K. Natarajan, J. Rundback, I.R. Sadiq, S.K. Liu, N. Bhalla, M.L. Raja, B.S. Weinstock, J. Cynamon, F.F. Elmasri, M.J. Garcia, M. Kumar, J. Ayerdi, P. Soukas, W. Kuo, P.-Y. Liu, S.Z. Goldhaber, S.I. Investigators, A Prospective, Single-Arm, Multicenter Trial of Ultrasound-Facilitated, Catheter-Directed, Low-Dose Fibrinolysis for Acute Massive and Submassive Pulmonary Embolism The SEATTLE II Study, Jacc Cardiovasc Interventions 8 (2015) 1382–1392. https://doi.org/10.1016/j.jcin.2015.04.020.

**Supplementary Data**

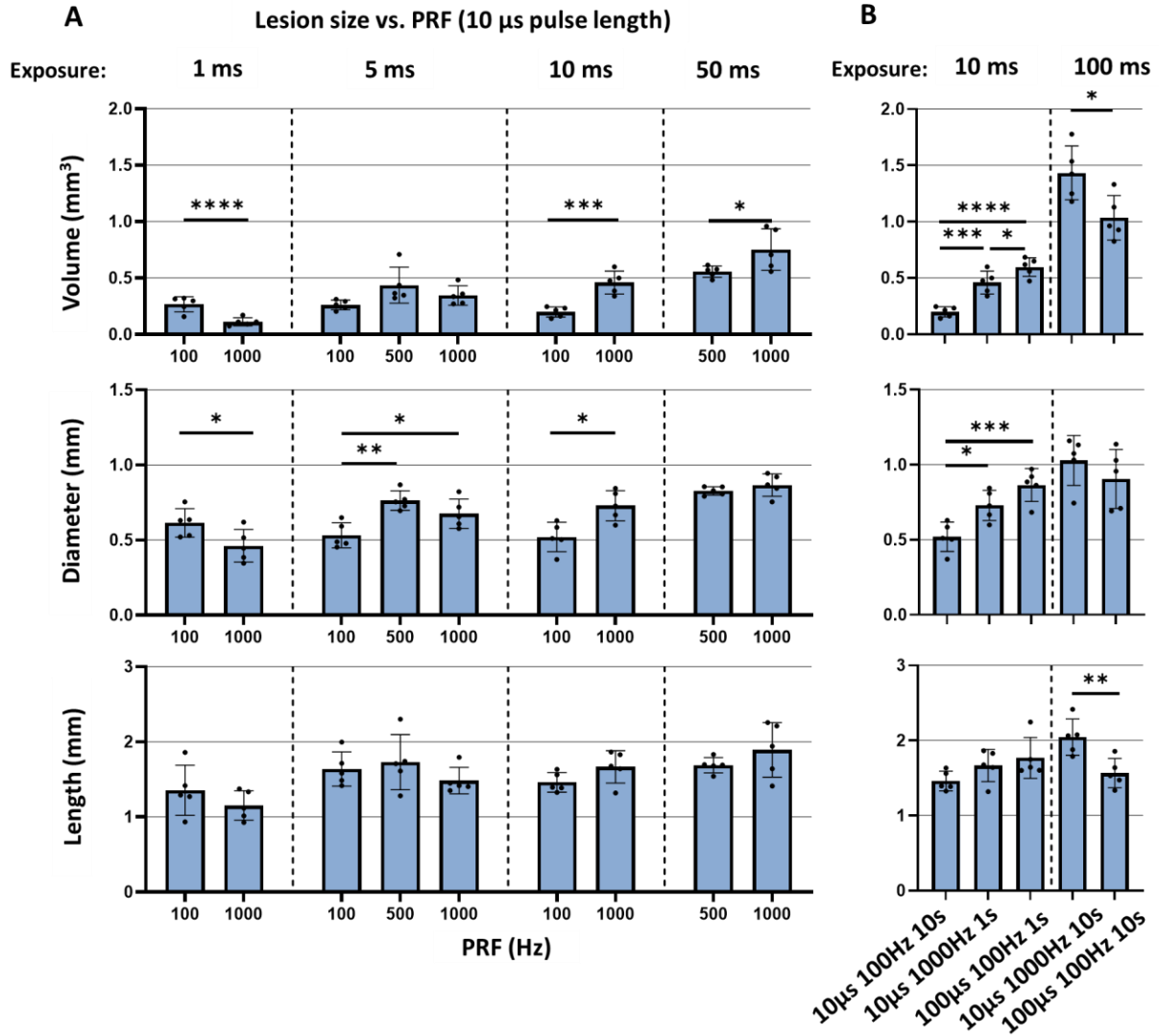

*Supplementary Figure 1: (A) Lesion measurements as a function of total exposure time and PRF for a pulse length of 10 μs. (B) Comparison of lesion measurements for pulsing schemes with the same ultrasound exposure durations.*

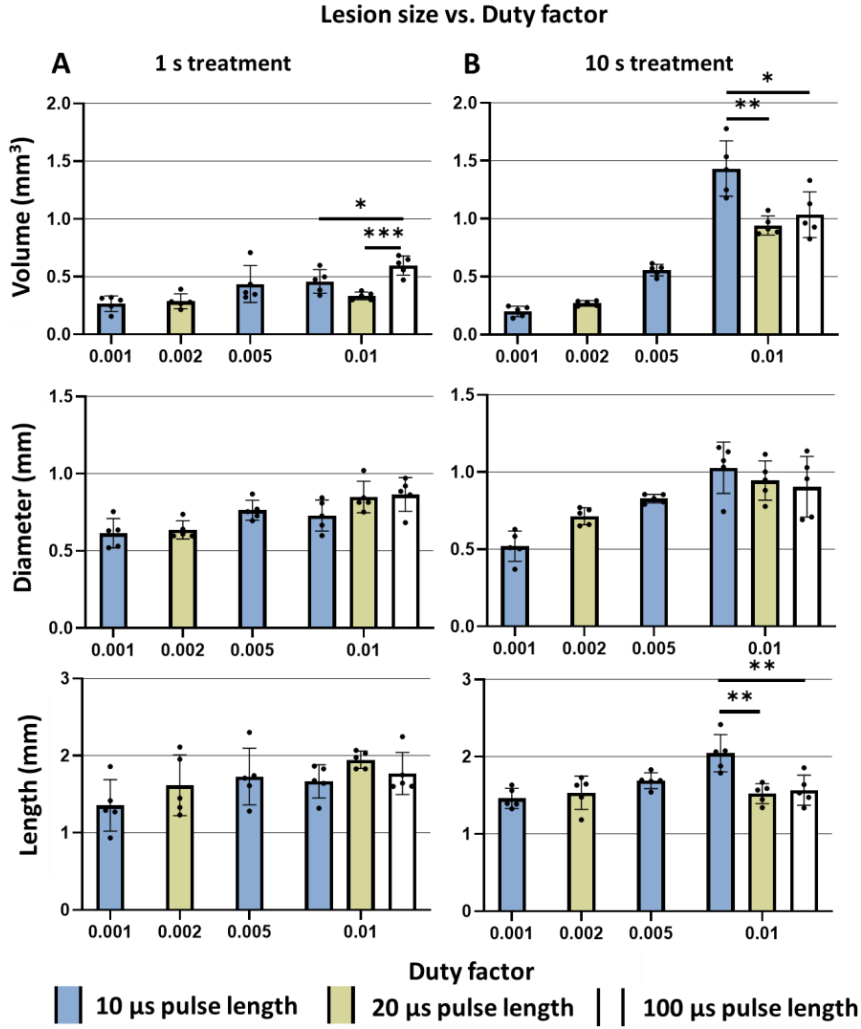

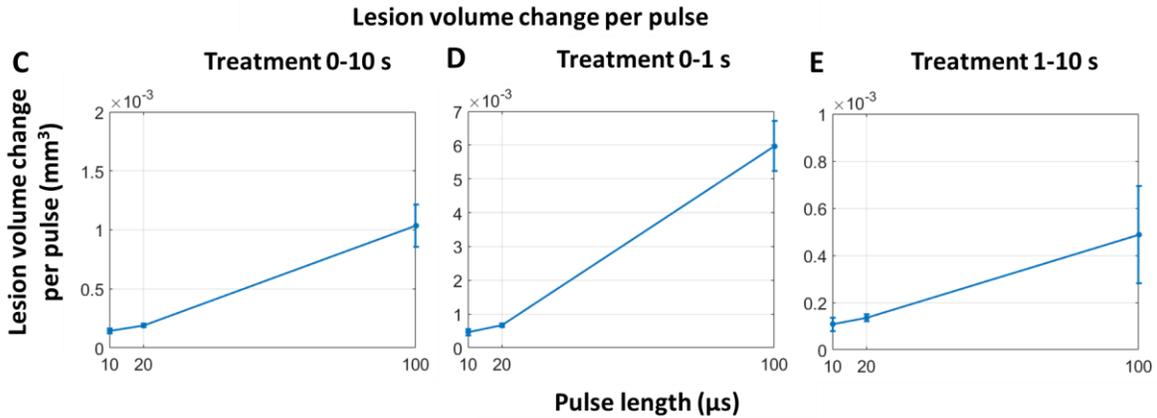

*Supplementary Figure 2: Lesion measurements (volume, diameter and length, A, B) as a function of duty factor for treatment durations of 1 and 10 seconds with pulse lengths of 10, 20 and 100 µs. Lesion volume change per pulse as a function of pulse length for different treatment durations of (C) 0–10 seconds, (D) 0–1 seconds, and (E) 1–10 seconds, obtained by subtracting the average*

*volume for 1 second duration treatments from the average volume of 10 second treatments and dividing by the net number of pulses over 9 seconds.*

**Table 1 Overview of pulsing schemes tested, with N=5 for each row**

| Number | Pulse length (µs) | PRF (Hz) | Duty factor | Number of pulses | Treatment duration (s) | Total exposure time (ms) |
|---|---|---|---|---|---|---|
| 1 | 10 | 100 | 0.001 | 100 | 1 | 1 |
| 2 | 10 | 100 | 0.001 | 500 | 5 | 5 |
| 3 | 10 | 100 | 0.001 | 1000 | 10 | 10 |
| 4 | 10 | 500 | 0.005 | 500 | 1 | 5 |
| 5 | 10 | 500 | 0.005 | 5000 | 10 | 50 |
| 6 | 10 | 1000 | 0.010 | 100 | 0.1 | 1 |
| 7 | 10 | 1000 | 0.010 | 200 | 0.2 | 2 |
| 8 | 10 | 1000 | 0.010 | 500 | 0.5 | 5 |
| 9 | 10 | 1000 | 0.010 | 1000 | 1 | 10 |
| 10 | 10 | 1000 | 0.010 | 2000 | 2 | 20 |
| 11 | 10 | 1000 | 0.010 | 5000 | 5 | 50 |
| 12 | 10 | 1000 | 0.010 | 10000 | 10 | 100 |
| 13 | 20 | 100 | 0.002 | 100 | 1 | 2 |
| 14 | 20 | 100 | 0.002 | 1000 | 10 | 20 |
| 15 | 20 | 500 | 0.010 | 500 | 1 | 10 |
| 16 | 20 | 500 | 0.010 | 5000 | 10 | 100 |
| 17 | 100 | 100 | 0.010 | 100 | 1 | 10 |
| 18 | 100 | 100 | 0.010 | 1000 | 10 | 100 |

Table 2 Statistical significance and fold change for lesion volumes for Figure 7

| 10 μs pulse, 1000 Hz PRF | | | | | | | |
|---|---|---|---|---|---|---|---|
| Treatment duration (s) | 0.1 | 0.2 | 0.5 | 1 | 2 | 5 | 10 |
| 0.1 | / | ns | ns | * | *** | **** | **** |
| 0.2 | 2.7 | / | ns | ns | ns | ** | **** |
| 0.5 | 3.1 | 1.1 | / | ns | ns | ** | **** |
| 1 | 4.1 | 1.5 | 1.3 | / | ns | ns | **** |
| 2 | 5.2 | 1.9 | 1.7 | 1.3 | / | ns | **** |
| 5 | 6.7 | 2.5 | 2.2 | 1.6 | 1.3 | / | *** |
| 10 | 10.9 | 4 | 3.5 | 2.6 | 2.1 | 1.6 | / |
| 10 μs pulse, 100 Hz PRF | | | | | | | |
| 1 | / | / | / | / | / | ns | ns |
| 5 | / | / | / | 1 | / | / | ns |
| 10 | / | / | / | 0.7 | / | 0.8 | / |

Fold change of the Mean
Significance

| Symbol | |
|---|---|
| ns | P>0.05 |
| * | P<0.05 |
| ** | P<0.01 |
| *** | P<0.001 |
| **** | P<0.0001 |

Table 3_1 Statistical significance and fold change for lesion volumes for Figure 8A

| 10 µs pulse, 1 s treatment | | | |
|---|---|---|---|
| PRF (Hz) | 100 | 500 | 1000 |
| 100 | / | ns | ns |
| 500 | 1.6 | / | ns |
| 1000 | 1.7 | 1.1 | / |
| 10 µs pulse, 5 s treatment | | | |
| 100 | / | | *** |
| 1000 | 2.9 | / | / |
| 10 µs pulse, 10 s treatment | | | |
| 100 | / | ** | **** |
| 500 | 2.8 | / | **** |
| 1000 | 7.2 | 2.6 | / |

| 10 µs pulse, 1000 Hz PRF | | | | | |
|---|---|---|---|---|---|
| Treament duration (s) | 0.1 | 0.5 | 1 | 5 | 10 |
| 0.1 | / | ns | * | **** | **** |
| 0.5 | 3.1 | / | ns | ** | **** |
| 1 | 4.1 | 1.3 | / | ns | **** |
| 5 | 6.7 | 2.2 | 1.6 | / | *** |
| 10 | 10.9 | 3.5 | 2.6 | 1.6 | / |
| 10 µs pulse, 500 Hz PRF | | | | | |
| 1 | / | / | / | / | ns |
| 10 | / | / | 1.3 | / | / |
| 10 µs pulse,100 Hz PRF | | | | | |
| 1 | / | / | / | ns | ns |
| 5 | / | / | 1 | / | ns |
| 10 | / | / | 0.7 | 0.8 | / |

Table 3_2 Statistical significance and fold change for lesion volumes for Figure 8B

| 10 µs pulse, 1 s treatment | | | |
|---|---|---|---|
| PRF (Hz) | 100 | 500 | 1000 |
| 100 | / | **** | **** |
| 500 | 0.3 | / | ns |
| 1000 | 0.2 | 0.5 | / |
| 10 µs pulse, 5 s treatment | | | |
| 100 | / | / | **** |
| 1000 | 0.3 | / | / |
| 10 µs pulse, 10 s treatment | | | |
| 100 | / | ** | * |
| 500 | 0.6 | / | ns |
| 1000 | 0.7 | 1.3 | / |

| Fold change of the Mean |
|---|
| Significance |

| Symbol | |
|---|---|
| ns | P>0.05 |
| * | P<0.05 |
| ** | P<0.01 |
| *** | P<0.001 |
| **** | P<0.0001 |

**Table 4_1 Statistical significance and fold change for lesion volumes for Figure 9A**

| 1 s treatment, 100 Hz PRF | | | | 1 s treatment, 10 µs pulse | | | |
|---|---|---|---|---|---|---|---|
| Pulse length (µs) | 10 | 20 | 100 | PRF (Hz) | 100 | 500 | 1000 |
| 10 | / | ns | **** | 100 | / | ns | ns |
| 20 | 1.1 | / | **** | 500 | 1.6 | / | ns |
| 100 | 2.2 | 2.1 | / | 1000 | 1.7 | 1.1 | / |
| 10 s treatment, 100 Hz PRF | | | | 10 s treatment, 10 µs pulse | | | |
| 10 | / | ns | **** | 100 | / | ** | **** |
| 20 | 1.4 | / | **** | 500 | 2.8 | / | **** |
| 100 | 5.2 | 3.8 | / | 1000 | 7.2 | 2.6 | / |
| 1 s treatment, 500 Hz PRF | | | | 1 s treatment, 20 µs pulse | | | |
| 10 | / | ns | / | 100 | / | ns | / |
| 20 | 0.8 | / | / | 500 | 1.2 | / | / |
| 10 s treatment, 500 Hz PRF | | | | 10 s treatment, 20 µs pulse | | | |
| 10 | / | **** | / | 100 | / | ns | / |
| 20 | 1.7 | / | / | 500 | 3.5 | / | / |

**Table 4_2 Statistical significance and fold change for lesion volumes for Figure 9B**

| 1 s treatment, 100 Hz PRF | | | |
|---|---|---|---|
| Pulse length (µs) | 10 | 20 | 100 |
| 10 | / | ns | **** |
| 20 | 1.1 | / | **** |
| 100 | 2.2 | 2.1 | / |
| 10 s treatment, 100 Hz PRF | | | |
| 10 | / | ns | **** |
| 20 | 1.4 | / | **** |
| 100 | 5.2 | 3.8 | / |
| 1 s treatment, 500 Hz PRF | | | |
| 10 | / | ns | / |
| 20 | 0.8 | / | / |
| 10 s treatment, 500 Hz PRF | | | |
| 10 | / | **** | / |
| 20 | 1.7 | / | / |

| | |
|---|---|
| Fold change of the Mean | |
| Significance | |

| Symbol | |
|---|---|
| ns | P>0.05 |
| * | P<0.05 |
| ** | P<0.01 |
| *** | P<0.001 |
| **** | P<0.0001 |

**Table 5_1 Statistical significance and fold change for lesion volumes for supplementary Figure 1A**

| 1 ms expoure, 10 μs pulse | | | |
|---|---|---|---|
| PRF (Hz) | 100 | 500 | 1000 |
| 100 | / | / | ** |
| 1000 | 0.4 | / | / |
| 5 ms expoure, 10 μs pulse | | | |
| 100 | / | ns | ns |
| 500 | 1.7 | / | ns |
| 1000 | 1.3 | 0.8 | / |
| 10 ms expoure, 10 μs pulse | | | |
| 100 | / | / | *** |
| 1000 | 2.3 | / | / |
| 50 ms expoure, 10 μs pulse | | | |
| 500 | / | * | / |
| 1000 | 1.4 | / | / |

| 10 μs pulse, 100 Hz PRF | | | | |
|---|---|---|---|---|
| Exposure (ms) | 1 | 5 | 10 | 50 |
| 1 | / | ns | ns | / |
| 5 | 1 | / | ns | / |
| 10 | 0.7 | 0.8 | / | / |
| 10 μs pulse, 500 Hz PRF | | | | |
| 5 | / | / | / | ns |
| 50 | / | 1.3 | / | / |
| 10 μs pulse, 1000 Hz PRF | | | | |
| 1 | / | * | ** | **** |
| 5 | 3.1 | / | ns | *** |
| 10 | 4.1 | 1.3 | / | ** |
| 50 | 6.7 | 2.2 | 1.6 | / |

**Table 5_2 Statistical significance and fold change for supplementary Figure 1B**

| 10 ms exposure time | | | |
|---|---|---|---|
| | 10μs 100Hz 10s | 10μs 1000Hz 1s | 100μs 100Hz 1s |
| 10μs 100Hz 10s | / | *** | **** |
| 10μs 1000Hz 1s | 2.3 | / | * |
| 100μs 100Hz 1s | 3 | 1.3 | / |
| 100 ms exposure time | | | |
| 10μs 100Hz 10s | / | * | / |
| 10μs 1000Hz 1s | 0.7 | / | / |

| | |
|---|---|
| Fold change of the Mean | |
| Significance | |
| Symbol | |
| ns | P>0.05 |
| * | P<0.05 |
| ** | P<0.01 |
| *** | P<0.001 |
| **** | P<0.0001 |

**Table 6 Statistical significance and fold change for supplementary Figure 2A**

| 1 s treatment, 10 μs pulse | | | |
|---|---|---|---|
| Duty factor | 0.001 | 0.005 | 0.01 |
| 0.001 | / | ns | ns |
| 0.005 | 1.6 | / | ns |
| 0.01 | 1.7 | 1.1 | / |

| 10 s treatment, 10 μs pulse | | | |
|---|---|---|---|
| 0.001 | / | ** | **** |
| 0.005 | 2.8 | / | **** |
| 0.01 | 7.2 | 2.6 | / |

| 0.01 duty factor, 1 s treatment | | | |
|---|---|---|---|
| Pulse length (μs) | 10 | 20 | 100 |
| 10 | / | ns | * |
| 20 | 0.7 | / | *** |
| 100 | 1.3 | 1.8 | / |

| 0.01 duty factor, 10 s treatment | | | |
|---|---|---|---|
| 10 | / | ** | * |
| 20 | 0.7 | / | ns |
| 100 | 0.7 | 1.1 | / |

| 1 s treatment, 20 μs pulse | | |
|---|---|---|
| Duty factor | 0.002 | 0.01 |
| 0.002 | / | ns |
| 0.01 | 1.2 | / |

| 10 s treatment, 20 μs pulse | | |
|---|---|---|
| 0.002 | / | **** |
| 0.01 | 3.5 | / |

Fold change of the Mean
Significance

| Symbol | |
|---|---|
| ns | P>0.05 |
| * | P<0.05 |
| ** | P<0.01 |
| *** | P<0.001 |
| **** | P<0.0001 |